\documentclass[a4paper,fleqn,usenatbib]{mnras}

\usepackage{newtxtext,newtxmath}

\usepackage[T1]{fontenc}
\usepackage{ae,aecompl}

\usepackage{graphicx}	
\usepackage{amsmath}	
\usepackage{amssymb}	

\title[Mid-IR imaging of HR 8799]{New constraints on the HR 8799 planetary system from mid-infrared direct imaging \thanks{Based on observations collected at the European Southern Observatory under ESO programme 0101.C-0580(A).}}

\author[D. J. M. Petit dit de la Roche et al.]{D. J. M. Petit dit de la Roche,$^{1}$\thanks{E-mail: dominique.petit@eso.org}
    M. E. van den Ancker,$^{1}$
    M. Kissler-Patig,$^{2,3}$\newauthor
    V. D. Ivanov,$^{1}$
    D. Fedele$^{4}$
    \\
    $^{1}$European Southern Observatory, Karl-Schwartzschild-Strasse 2, 85748 Garching, Germany\\
    $^{2}$European Space Agency, Camino Bajo del Castillo, s/n., Urb. Villafranca del Castillo, 28692 Villanueva de la Ca\~{n}ada, Madrid, Spain\\
    $^{3}$Universit\"ats-Sternwarte, Ludwig-Maximilians-Universit\"at M\"unchen, Scheinerstr 1, D-81679 M\"unchen, Germany\\
    $^{4}$INAF-Osservatorio Astrofisico di Arcetri, Largo E. Fermi 5, 50125 Firenze, Italy
    }

\date{Accepted XXX. Received YYY; in original form ZZZ}

\pubyear{2019}

\begin{document}
\label{firstpage}
\pagerange{\pageref{firstpage}--\pageref{lastpage}}
\maketitle

\begin{abstract}
Direct imaging is a tried and tested method of detecting exoplanets in the near infrared, but has so far not been extended to longer wavelengths. New data at mid-IR wavelengths (8-20\,\textmu m) can provide additional constraints on planetary atmospheric models. We use the VISIR instrument on the VLT to detect or set stringent limits on the 8.7\,\textmu m flux of the four planets surrounding HR 8799, and to search for additional companions. We use a novel circularised PSF subtraction technique to reduce the stellar signal and obtain instrument limited background levels and obtain optimal flux limits. The BT SETTL isochrones are then used to determine the resulting mass limits. We find flux limits between 0.7 and 3.3\,mJy for the J8.9 flux of the different planets at better than $5\sigma$ level and derive a new mass limit of 30\,M\textsubscript{Jup} for any objects beyond 40\,AU. While this work has not detected planets in the HR 8799 system at 8.7\,\textmu m, it has found that an instrument with the sensitivity of VISIR is sufficient to detect at least 4 known hot planets around close stars, including $\beta$ Pictoris b (1700\,K, 19\,pc), with more than $5\sigma$ certainty in 10 hours of observing time in the mid-IR. 
\end{abstract}




\section{Introduction}

HR 8799 is a young A star located at $41.3\pm 0.2$\,pc \citep{Marois2008,Gaia2018} that is unique in that it has four directly imaged exoplanets surrounding it. The planets orbit between 17 and 68 AU and have masses between 7 and 10\,M\textsubscript{Jup} \citep{Marois2008, Marois2010}. The system also contains a warm dust belt within the innermost planet e and a Kuiper belt-like debris disk outside the planets orbits from 145\,AU to 450\,AU \citep{Su2009, Booth2016}. \citet{Booth2016} suggested a possible fifth planet of 1.25\,M\textsubscript{Jup} at 110\,AU that could be responsible for the gap between the outer planet b and the debris disk, although the mass could be lower if the planet is further out or on an eccentric orbit. Due to this unique position the planets have been studied extensively in imaging and spectroscopy across the near infrared (IR) from the J to the M band since their discovery \citep[e.g.][]{Barman2011, Galicher2011, Zurlo2016, Petit2018}. 

So far no direct imaging of planets has been done beyond 4.8\,\textmu m. Longer wavelength searches with \textit{Spitzer} and \textit{WISE} have revealed a number of brown dwarf companions at large separations and \citet{Geissler2008} have done ground based mid-infrared imaging of brown dwarfs in binary systems with VISIR, but no planets have been imaged \citep{Luhman2007, Luhman2012}. Mid-IR observations can provide further constraints of planetary atmospheric models and the mid-IR wavelength range also contains biosignatures, which could indicate the presence of biological processes \citep{Meadows2010, Rauer2011}. \textit{Spitzer} has made observations of secondary transits, occultations of exoplanets by their host stars, at longer wavelengths \citep[e.g.][]{Deming2005, Charbonneau2005, Deming2007}, resulting in light curves and broad band emission spectra of some transiting exoplanets. These are useful for comparison to imaging data, but the planets observed with \textit{Spitzer} cover a different area of parameter space than directly imaged planets due to the limitations of both methods. Finally, ground-based detection of exoplanets in the mid-IR will help the analysis and interpretation of future JWST exoplanet observations.

We aim to expand current observations to cover the mid-IR wavelength regime by imaging the HR 8799 system with the VLT Imager and Spectrometer for the mid-IR (VISIR, \citealt{lagage2004}) at 8.7\,\textmu m. Additionally, we set constraints on potentially undiscovered companions further out from the star. While both flux limits and mass limits for additional companions have been calculated before in near IR bands \citep{Metchev2009, Serabyn2010, Currie2011, Galicher2011, Esposito2013, Zurlo2016}, only \citet{Metchev2009} and \citet{Serabyn2010} have searched for companions beyond 2" and then only up to 3.5" and 4" respectively. VISIR will allow us to expand the search to a factor two larger separations, covering new sky areas. 

The data and analysis are described in Section \ref{sec:observations}. The final images and derived limits are presented in Section \ref{sec:results} and we present our conclusions in Section \ref{sec:conclusion}. 

\section{Observations and data analysis}
\label{sec:observations}
The observations of HR 8799 were taken with VISIR on the VLT UT3 telescope in the small-field imaging mode with a plate scale of 45\,mas\,px$^{-1}$. They were taken in the J8.9 filter ($\lambda_0=8.72$\,\textmu m, $\Delta \lambda = 0.73$\,\textmu m) using the pupil tracking mode, in which field rotation is enabled to achieve a more stable image. The chopping and nodding sequence were enabled to subtract sky background with a chop throw of 8" and a chopping frequency of 4\,Hz. The nodding direction was perpendicular to the chop direction. Due to the chop throw, the total usable field of view was slightly smaller than 16"x16". The observations were carried out between August and October of 2018, with a total on-target integration time of 8.7\,h as shown in Table~\ref{tab:observations}.



\begin{table}
\caption{\label{tab:observations} A summary of the VISIR observations of HR 8799 taken between August and October of 2018. The integration time refers to the total on-source integration time of all the data taken on each night. The total on-target time over all nights is provided in the bottom line. The image quality is determined by the full width half maximum (FWHM) of the PSF during the night. }
\centering
\begin{tabular}{cccc}
\hline
Date       & \begin{tabular}[c]{@{}c@{}}Integration\\ time\end{tabular} & \begin{tabular}[c]{@{}c@{}} Airmass \end{tabular} & \begin{tabular}[c]{@{}c@{}}Image quality\\ FWHM \end{tabular}  \\ \hline
23-08-2018  & 1h & 1.5 & 0.26" \\ 
09-09-2018   & 1h & 1.5 & 0.24"\\ 
14-09-2018  & 1h & 1.6 & 0.26"\\ 
16-09-2018  & 0.4h & 1.4 & 0.29"\\
03-10-2018  & 1h & 1.5 & 0.24" \\
11-10-2018 & 1.3h & 1.6 & 0.31"\\
12-10-2018 & 1h & 1.5 & 0.27"\\
16-10-2018 & 1h & 1.5 & 0.32"\\
17-10-2018 & 1h & 1.5 & 0.29"\\ 
Total & 8.7h & & \\\hline
\end{tabular}

\end{table}



Data are provided in the form of time averaged chop difference frames with integration times of 50\,s and units of counts per detector integration time (DIT, 0.0114\,s). Since the automated VISIR data reduction\footnote{\url{https://www.eso.org/sci/software/pipelines/visir/visir-pipe-recipes.html}} is not equipped to reduce pupil stabilised data, the reduction was performed with special-purpose Python scripts. Images were pairwise subtracted between different nod positions to reduce non-common path errors. Beam combination and centering was achieved through fitting Gaussian functions to each of the sources in the nod difference images. The resulting images were stacked into cubes for each night.

\begin{figure}
    \centering
    \includegraphics[width=\columnwidth]{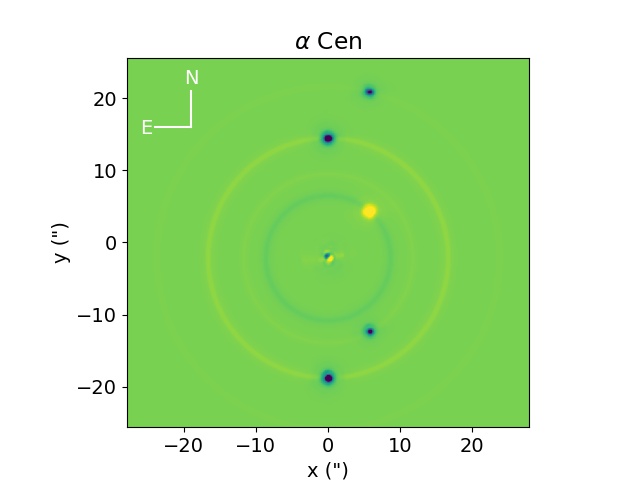}
    \caption{Example of circularised PSF subtraction for the binary stars of $\alpha$ Cen. $\alpha$ Cen A is the central source and $\alpha$ Cen B the companion. The central source is subtracted, while the subtraction of the circularised PSF results in a dark ring at the separation of the second source. The second source is still clearly visible, as are in this case the chop/nod shadows of both sources and their corresponding bright rings. }
    \label{fig:aCen}
\end{figure}

Traditional angular differential imaging (ADI, \citealt{Marois2006}), requires enough time to have passed between the science and reference images for the planet to have moved by at least $1.2\lambda/D$ to avoid self-subtraction of the planetary point spread function (PSF). However, due to the small angular separations of the inner planets d and e this time is sufficiently large that there are (almost) no reference images available for most of the data. Instead, we apply a novel circularised PSF subtraction technique. A circularised PSF of the science data was created by rotating it 1 to 360 degrees in 1 degree steps and averaging over all rotated images. The resulting PSF is a circularly symmetric version of the science data that can then be subtracted from the original to reduce stellar contributions. Any visible secondary source will show up in this circularised PSF as a ring around the central star. The selfsubtraction caused by this ring is expected to be minimal (at most 5\% for the innermost planet and less for the planets further out) due to averaging of the source brightness over the full 360 degrees of rotation. An example of the method for the $\alpha$ Cen system is shown in Fig.~\ref{fig:aCen}. Circularised PSF subtraction has the advantage that, unlike traditional ADI, it is not limited to sources with sufficient field rotation and can therefore be used more widely and on shorter observations. For instruments such as VISIR it is also less sensitive to variations in atmospheric conditions than traditional PSF subtraction, since in this case simultaneous observation of the PSF is not possible. 

Once the stellar component has been reduced the images are derotated such that north is up and then the derotated images are combined into a single master image using a weighted sigma-clipped median function with a threshold of $3\sigma$. The weights were determined by the standard deviation and thus the quality of each image. The master image has a total integration time of 8.7 hours and therefore an expected $5\sigma$ background sensitivity of 0.7\,mJy. The master image was calibrated by comparing it to an image of the stacked and derotated data where no PSF had been subtracted. HR 8799 is expected to have a flux density of ~430\,mJy in the J8.9 filterband. The conversion factor of the data in ADU\,/\,Jy was calculated following the procedures for reference stars in the VISIR pipeline manual and applied to the reduced master image. 

\begin{figure*}
    \centering
    \includegraphics[width = \textwidth]{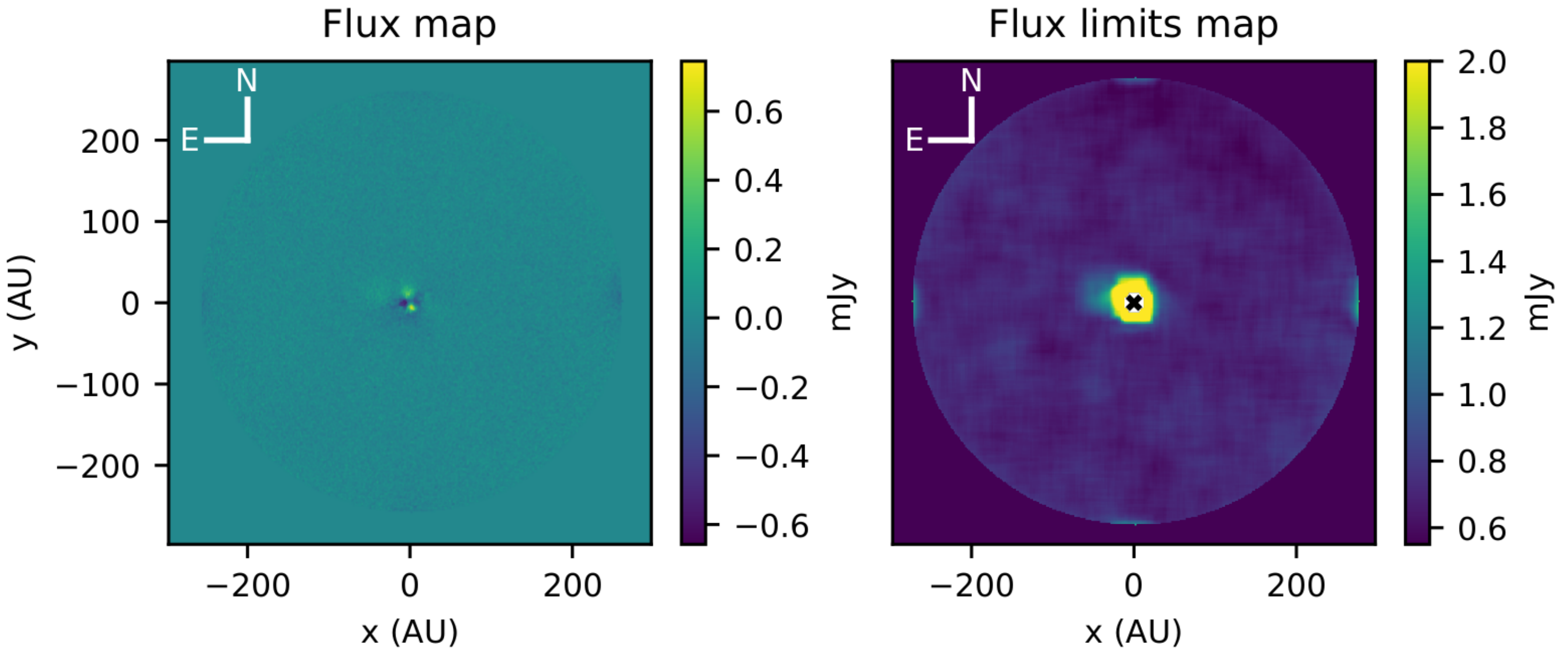}
    \caption{\emph{Left:} Reduced master image of HR 8799, centered on the star. The planets are not visible in the reduced data. Positive and negative structures are visible in the center from imperfect subtraction of the stellar PSF. \emph{Right:} $5\sigma$ flux limit map. The center of the image around the star where the limits are highest has been occulted for clarity. The decrease of the limits with radius shows that stellar features dominate the inner 40\,AU, and further out a background sensitivity is achieved.  }
    \label{fig:maps}
\end{figure*}

\section{Results and Discussion}
\label{sec:results}

Fig.~\ref{fig:maps} shows the reduced master image and the map of the $5\sigma$ flux limits derived from the master image. While none of the planets are detected, upper limits can be placed on their emission, as shown in Table~\ref{tab:limits}. The upper limit of the flux was determined by calculating the expected flux in a Gaussian function with a full-width-half-maximum of one resolution element of the telescope and a peak flux of $5\sigma$ of the background in the surrounding area. The limits found for HR 8799 b and c correspond to the expected background sensitivity of the instrument over the observed time. The higher limits on HR 8799 d and e are the result of imperfectly subtracted stellar residuals, as due to seeing the stellar PSFs in the science data were no longer entirely circularly symmetrical. 

The right panel in Fig.~\ref{fig:maps} shows that the $5\sigma$ flux limits decrease with distance from the star in all directions, further supporting that the increased limits are due to stellar residuals. No other structures are visible and the background sensitivity is achieved beyond \textasciitilde 40\,AU. 

\begin{figure}
    \centering
    \includegraphics[width=0.5\textwidth]{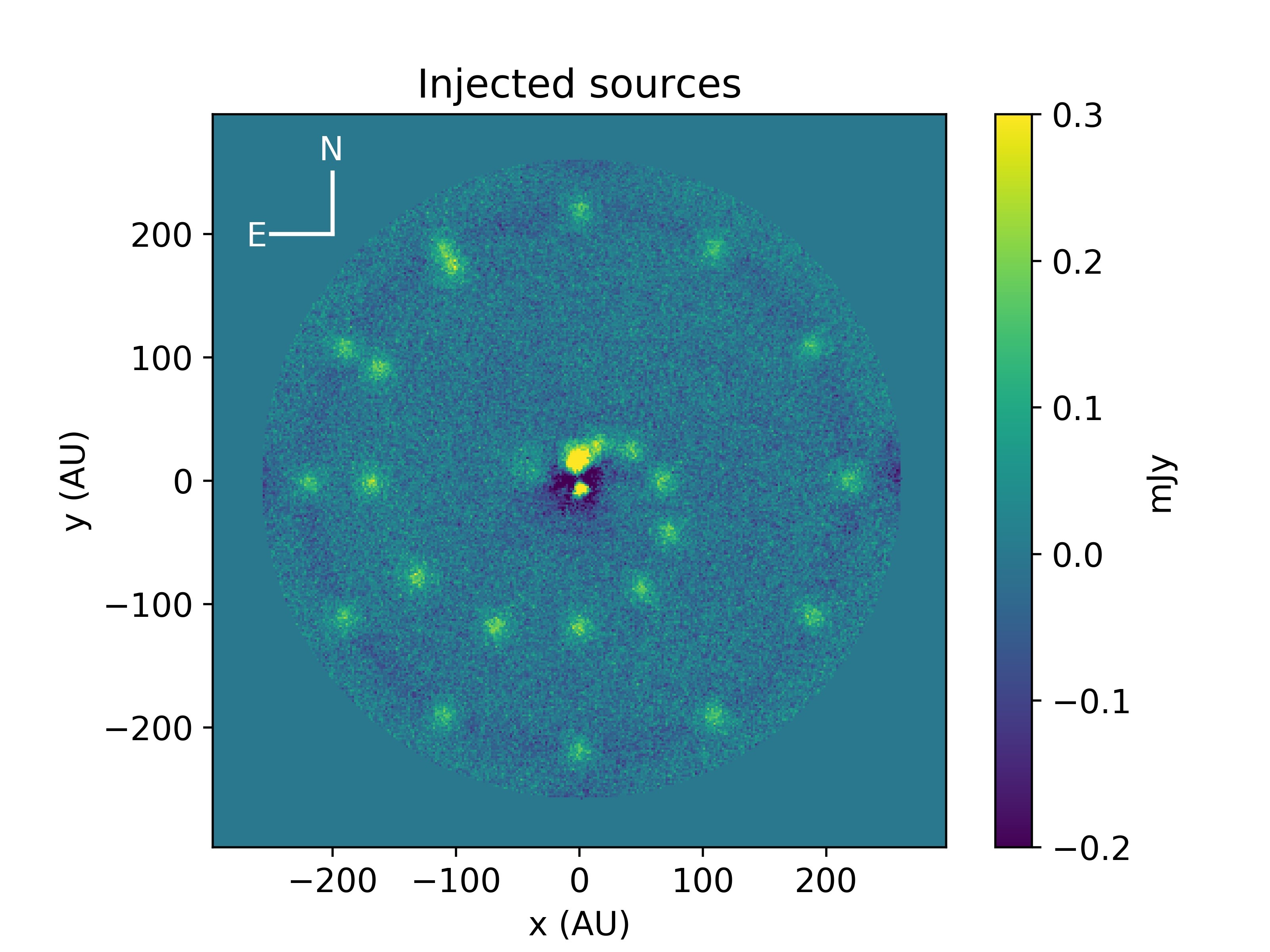}
    \caption{Reduced master image of HR 8799 with injected sources at the 5\,$\sigma$ limit. 12 sources are injected at increasing radius and 12 at an outer radius of 200\,AU. The sources in the spiral are retrieved at 4.5\,$\sigma$ and sources in the ring at 3.5$\,\sigma$. The lower certainty retrievals are due to the sources contributing to the PSF that is subtracted from the background.}
    \label{fig:inj}
\end{figure}

To verify these limits we injected 24 fake sources into the data before the psf subtraction. Figure \ref{fig:inj} shows the reduced data with the injected sources. 12 sources are injected at increasing radii and 12 are injected at the same radius of 200\,AU. Each source has a flux equal to the 5\,$\sigma$ flux limit at that radius. The sources in the spiral are retrieved at 4.5$\,\sigma$ confidence, but the sources on the ring at only 3.5$\,\sigma$. In both cases the difference with the injected magnitude is due to the contribution of the injected sources to the circularised PSF. For the spiral there is only one source at each radius and the effect is small, but for the ring there are 12 sources contributing to the PSF at 200\,AU, resulting in a brighter ring being subtracted. This results in a dark ring at that radius and a reduced magnitude of the retrieved sources. Despite this effect all injected sources are retrieved and since planets are not expected to have identical orbital separations within one system, we conclude that our limits are valid. 

\begin{figure}
    \centering
    \includegraphics[width = \columnwidth]{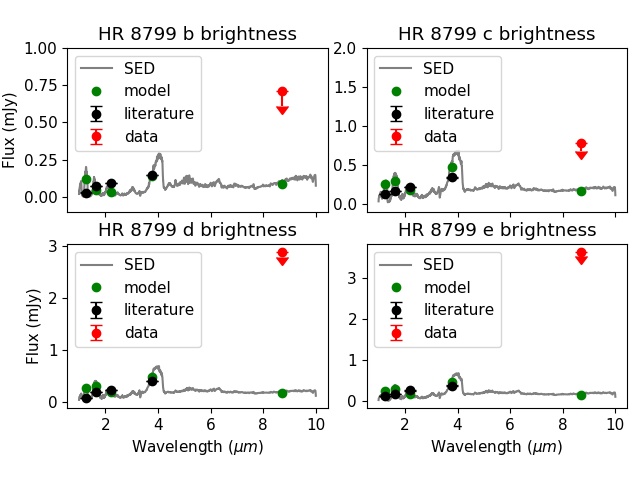}
    \caption{Theoretical spectral energy distributions (SED, grey) for HR 8799 bcde from top left to bottom right using the BT SETTL model with temperatures and surface gravities from \citet{Marois2008, Marois2010}. The flux density values of the earlier measurements are marked in black and the values derived from the model are marked in green. The $5\sigma$ upper limits calculated in this work are marked in red. All four planets have expected J8.9 flux densities below the derived limits, resulting in a nondetection of the planets in the data. }
    \label{fig:seds}
\end{figure}

Fig.~\ref{fig:seds} shows the model fits to the near infrared spectral energy distributions of the four planets \citep{Marois2008, Marois2010}. Our flux limits and the expected model predicted flux densities of each planet in the J8.9 band are marked as well. This shows that the expected flux density lies below the derived limits in all cases. 

\begin{figure}
    \centering
    \includegraphics[scale=.4]{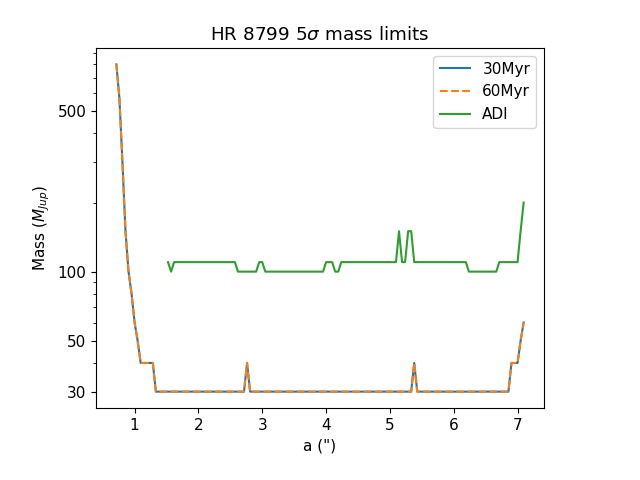}
    \caption{Upper mass limits ($5\sigma$) for the mass of any objects in the HR 8799 system as a function of separation from the star. At small separations the flux from stellar residuals causes high mass limits, but at 1" (40\,AU) these become negligible and the background dominates for the circularised PSF subtracted data. The mass limit levels out to 30\,M\textsubscript{Jup}, as indicated by the black dotted line. The increase furthest out is due to the chop/nod shadows located at 8". The mass limits resulting from data reduced with traditional ADI beyond 1.5" are shown in green and are around 100\,M\textsubscript{Jup}.}
    \label{fig:mass_limits}
\end{figure}

To convert the obtained flux limits into mass limits, we condensed the flux limit map from Fig.~\ref{fig:maps} into a contrast curve, which is then converted into mass limits for any objects in the system. This is done using the BT SETTL evolutionary models of stars, brown dwarfs and exoplanets \citep{Allard2012} and assuming an age for the system. Two ages have been considered in previous works: 30\,Myr and 60\,Myr, depending on whether the authors believe HR 8799 to be part of the Columba local association (eg. \citealt{Galicher2011, Zurlo2016}) or not \citep{Metchev2009, Hinz2010}. \citet{Doyon2010} and \citet{Zuckerman2011} have determined that HR 8799 is likely to belong to the Columba association and as such has an age of 30\,Myr, while \citet{Hinz2010} have not found this to be the case and used an age of 60\,Myr based on the original considerations regarding the disk mass, stellar class, HR diagram location and galactic motion by \citet{Marois2008}. For this work we have examined the mass limits at both ages and found them to be nearly identical. The difference in the ages means a difference in available cooling time, resulting in $\sim$150\,K difference between the two models at $\sim$30\,M\textsubscript{Jup}. The resulting difference in 8.7\,\textmu m flux is less than 0.1\,mJy with the result that the returned mass limits are very similar. Our results are shown in Fig.~\ref{fig:mass_limits} and show a sharp increase in mass inward of 40\,AU, where stellar wings dominate the background. Outside of this distance the line levels out at 30\,M\textsubscript{Jup}, which corresponds to the sensitivity limits of the instrument. Our result is in agreement with earlier work by \citet{Marois2008, Marois2010} and \citet{Wang2018b} which places the masses of the four planets between 7 and 10 M\textsubscript{Jup}. The calculated limits are further in agreement with limits set in the near infrared within 2" by \citet{Zurlo2016}. The sharp increase at around 8" is due to the chop/nod shadows of the source in the data. The steps in the mass limits are caused by the stepsize in mass in the isochrone models, which is 10\,M\textsubscript{Jup} between planet masses of 20 and 100\,M\textsubscript{Jup}. 

Mass limits for the same data reduced with ADI are also shown in Fig.~\ref{fig:mass_limits} in green. Here an age of 30\,Myr is assumed. The mass limits retrieved with this technique are around 100\,M\textsubscript{Jup}, three times higher than for the circularised PSF subtraction. This difference in mass limits represents a difference in flux limits of 0.26\,mJy. Limits inside 1.5" are not shown due to insufficient field rotation closer to the star. 


\begin{table}
\centering
\caption{\label{tab:limits} Detection limits for the four planets in the HR 8799 system. Separations and position angles are taken from \citet{Marois2008,Marois2010} with position angles corrected for orbital motion based on the planets periods.}

\begin{tabular}{cccc}
\hline
Planet & Separation & \begin{tabular}[c]{@{}c@{}}Position\\ angle\end{tabular} &  \begin{tabular}[c]{@{}c@{}}$5\sigma$ limits\\ (mJy)\end{tabular} \\ \hline
HR 8799 b & 1.7" & 71$^\circ$     & <0.7         \\
HR 8799 c & 1.0" & 332$^\circ$     & <0.8         \\
HR 8799 d & 0.6" & 230$^\circ$     & <2.9            \\
HR 8799 e & 0.4" & 292$^\circ$     & <3.3           \\ \hline
\end{tabular}

\end{table}

\section{Conclusions}
\label{sec:conclusion}
This work presents the first mid infrared direct imaging observations of the HR 8799 planetary system and places constraints on the fluxes and masses of objects between 40 and 330\,AU. It thus provides the most stringent limits obtained to date at the furthest separations so far. The results exclude further companions with fluxes of more than 0.7\,mJy or masses exceding 30\,M\textsubscript{Jup} with a $5\sigma$ certainty, given an age of 60\,Myr. 

The achieved flux limits were insufficient to detect the HR 8799 planets, but we demonstrated that VISIR can reach sufficient sensitivity to detect planets in other systems which are hotter and/or closer to Earth. Excluding the HR 8799 system, the exoplanet.eu database \citep{exoplaneteu, Schneider2011} contains 22 directly imaged planets with a listed temperature and a known surface gravity within 50\,pc. Of these, 4 planets are sufficiently bright to be imaged by instruments with similar sensitivities to VISIR at a $5\sigma$ detection level in less than 10 hours: these planets are $\beta$ Pic b, CD-35 2722 b, HD 116434 b and G196-3 b. $\beta$ Pic b has a temperature of 1800\,K and is located at a distance of 19.8\,pc. \citep{Chilcote2017, Gaia2018} It is expected to have an 8.7\,\textmu m flux density of 3.0\,mJy. While $\beta$ Pic b is currently very close to its star, \citet{Dupuy2019} have calculated that by 2028 the angular separation should be around 0.68-0.75" for a predicted eccentricity of 0.24. The flux limit is then 2.4\,mJy for 10 hours of observation and the planet becomes detectable. CD-35 2722 b and G196-3 b have similar temperatures and distances \citep{Wahhaj2011, Rebolo1998}, while HD 116434 b is a cooler, closer in planet with a temperature of 1300\,K at 11\,pc. \citep{Chauvin2017} 

Additionally, VISIR is being upgraded into the NEAR (New Earths in the Alpha Cen Region) instrument, which, thanks to the addition of adaptive optics, is reported to have a $10\sigma$ sensitivity of 0.9\,mJy in 1 hour, an improvement of approximately a factor of four. For the foreseeable future NEAR will only be observing the $\alpha$\,Cen system, but this kind of advancement in technology will allow the previously mentioned planets, as well as four additional planets, to be imaged in under an hour. Eleven more planets, including HR 8799 c and d, become accessible with up to 10 hours of observation time, demonstrating the potential of directly imaging exoplanets at mid-IR wavelengths with present-day facilities.  






\bibliographystyle{mnras}
\bibliography{bibliography} 

\begin{thebibliography}{}
\makeatletter
\relax
\def\mn@urlcharsother{\let\do\@makeother \do\$\do\&\do\#\do\^\do\_\do\%\do\~}
\def\mn@doi{\begingroup\mn@urlcharsother \@ifnextchar [ {\mn@doi@}
  {\mn@doi@[]}}
\def\mn@doi@[#1]#2{\def\@tempa{#1}\ifx\@tempa\@empty \href
  {http://dx.doi.org/#2} {doi:#2}\else \href {http://dx.doi.org/#2} {#1}\fi
  \endgroup}
\def\mn@eprint#1#2{\mn@eprint@#1:#2::\@nil}
\def\mn@eprint@arXiv#1{\href {http://arxiv.org/abs/#1} {{\tt arXiv:#1}}}
\def\mn@eprint@dblp#1{\href {http://dblp.uni-trier.de/rec/bibtex/#1.xml}
  {dblp:#1}}
\def\mn@eprint@#1:#2:#3:#4\@nil{\def\@tempa {#1}\def\@tempb {#2}\def\@tempc
  {#3}\ifx \@tempc \@empty \let \@tempc \@tempb \let \@tempb \@tempa \fi \ifx
  \@tempb \@empty \def\@tempb {arXiv}\fi \@ifundefined
  {mn@eprint@\@tempb}{\@tempb:\@tempc}{\expandafter \expandafter \csname
  mn@eprint@\@tempb\endcsname \expandafter{\@tempc}}}

\bibitem[\protect\citeauthoryear{{Allard}, {Homeier}  \& {Freytag}}{{Allard}
  et~al.}{2012}]{Allard2012}
{Allard} F.,  {Homeier} D.,   {Freytag} B.,  2012, \mn@doi [Philosophical
  Transactions of the Royal Society of London Series A]
  {10.1098/rsta.2011.0269}, \href
  {https://ui.adsabs.harvard.edu/abs/2012RSPTA.370.2765A} {370, 2765}

\bibitem[\protect\citeauthoryear{Barman, Macintosh, Konopacky  \&
  Marois}{Barman et~al.}{2011}]{Barman2011}
Barman T.~S.,  Macintosh B.,  Konopacky Q.~M.,   Marois C.,  2011, The
  Astrophysical Journal, 733, 65

\bibitem[\protect\citeauthoryear{Booth et~al.,}{Booth et~al.}{2016}]{Booth2016}
Booth M.,  et~al., 2016, \mn@doi [Monthly Notices of the Royal Astronomical
  Society: Letters] {10.1093/mnrasl/slw040}, 460, L10

\bibitem[\protect\citeauthoryear{Charbonneau et~al.,}{Charbonneau
  et~al.}{2005}]{Charbonneau2005}
Charbonneau D.,  et~al., 2005, The Astrophysical Journal, 626, 523

\bibitem[\protect\citeauthoryear{Chauvin et~al.}{Chauvin
  et~al.}{2017}]{Chauvin2017}
Chauvin G.,  et~al., 2017, \mn@doi [A\&A] {10.1051/0004-6361/201731152}, 605,
  L9

\bibitem[\protect\citeauthoryear{Chilcote et~al.}{Chilcote
  et~al.}{2017}]{Chilcote2017}
Chilcote J.,  et~al., 2017, \mn@doi [The Astronomical Journal]
  {10.3847/1538-3881/aa63e9}, 153, 182

\bibitem[\protect\citeauthoryear{{Currie} et~al.,}{{Currie}
  et~al.}{2011}]{Currie2011}
{Currie} T.,  et~al., 2011, \mn@doi [\apj] {10.1088/0004-637X/729/2/128}, \href
  {https://ui.adsabs.harvard.edu/abs/2011ApJ...729..128C} {729, 128}

\bibitem[\protect\citeauthoryear{Deming, Seager, Richardson  \&
  Harrington}{Deming et~al.}{2005}]{Deming2005}
Deming D.,  Seager S.,  Richardson L.~J.,   Harrington J.,  2005, \mn@doi
  [Nature] {10.1038/nature03507}, 434, 740

\bibitem[\protect\citeauthoryear{Deming, Harrington, Laughlin, Seager, Navarro,
  Bowman  \& Horning}{Deming et~al.}{2007}]{Deming2007}
Deming D.,  Harrington J.,  Laughlin G.,  Seager S.,  Navarro S.~B.,  Bowman
  W.~C.,   Horning K.,  2007, \mn@doi [The Astrophysical Journal]
  {10.1086/522496}, 667, L199

\bibitem[\protect\citeauthoryear{{Doyon}, {Lafreni{\`e}re}, {Artigau}, {Malo}
  \& {Marois}}{{Doyon} et~al.}{2010}]{Doyon2010}
{Doyon} R.,  {Lafreni{\`e}re} D.,  {Artigau} E.,  {Malo} L.,   {Marois} C.,
  2010, in In the Spirit of Lyot 2010. p.~E42

\bibitem[\protect\citeauthoryear{Dupuy, Brandt, Kratter  \& Bowler}{Dupuy
  et~al.}{2019}]{Dupuy2019}
Dupuy T.~J.,  Brandt T.~D.,  Kratter K.~M.,   Bowler B.~P.,  2019, \mn@doi [The
  Astrophysical Journal] {10.3847/2041-8213/aafb31}, 871, L4

\bibitem[\protect\citeauthoryear{Esposito et~al.}{Esposito
  et~al.}{2013}]{Esposito2013}
Esposito S.,  et~al., 2013, Astronomy \& Astrophysics, 549, A52

\bibitem[\protect\citeauthoryear{{Exoplanet Team}}{{Exoplanet
  Team}}{1995}]{exoplaneteu}
{Exoplanet Team} 1995, The Extrasolar Planet Encyclopedia

\bibitem[\protect\citeauthoryear{{Gaia Collaboration}}{{Gaia
  Collaboration}}{2018}]{Gaia2018}
{Gaia Collaboration} 2018, VizieR Online Data Catalog, 1345

\bibitem[\protect\citeauthoryear{Galicher, Marois, Macintosh, Barman  \&
  Konopacky}{Galicher et~al.}{2011}]{Galicher2011}
Galicher R.,  Marois C.,  Macintosh B.,  Barman T.,   Konopacky Q.,  2011,
  \mn@doi [The Astrophysical Journal] {10.1088/2041-8205/739/2/l41}, 739, L41

\bibitem[\protect\citeauthoryear{{Gei{\ss}ler}, {Chauvin}  \&
  {Sterzik}}{{Gei{\ss}ler} et~al.}{2008}]{Geissler2008}
{Gei{\ss}ler} K.,  {Chauvin} G.,   {Sterzik} M.~F.,  2008, \mn@doi [\aap]
  {10.1051/0004-6361:20078229}, \href
  {https://ui.adsabs.harvard.edu/abs/2008A&A...480..193G} {480, 193}

\bibitem[\protect\citeauthoryear{Hinz et~al.}{Hinz et~al.}{2010}]{Hinz2010}
Hinz P.~M.,  et~al., 2010, The Astrophysical Journal, 716, 417

\bibitem[\protect\citeauthoryear{{Lagage} et~al.}{{Lagage}
  et~al.}{2004}]{lagage2004}
{Lagage} P.~O.,  et~al., 2004, The Messenger, 117, 12

\bibitem[\protect\citeauthoryear{Luhman et~al.,}{Luhman
  et~al.}{2007}]{Luhman2007}
Luhman K.~L.,  et~al., 2007, \mn@doi [The Astrophysical Journal]
  {10.1086/509073}, 654, 570

\bibitem[\protect\citeauthoryear{Luhman et~al.,}{Luhman
  et~al.}{2012}]{Luhman2012}
Luhman K.~L.,  et~al., 2012, \mn@doi [The Astrophysical Journal]
  {10.1088/0004-637x/760/2/152}, 760, 152

\bibitem[\protect\citeauthoryear{Marois, Lafreni\`{e}re, Doyon, Macintosh  \&
  Nadeau}{Marois et~al.}{2006}]{Marois2006}
Marois C.,  Lafreni\`{e}re D.,  Doyon R.,  Macintosh B.,   Nadeau D.,  2006,
  The Astrophysical Journal, 641, 556

\bibitem[\protect\citeauthoryear{Marois et~al.}{Marois
  et~al.}{2008}]{Marois2008}
Marois C.,  et~al., 2008, \mn@doi [Science] {10.1126/science.1166585}, 322,
  1348

\bibitem[\protect\citeauthoryear{Marois, Zuckerman, Konopacky, Macintosh  \&
  Barman}{Marois et~al.}{2010}]{Marois2010}
Marois C.,  Zuckerman B.,  Konopacky Q.~M.,  Macintosh B.,   Barman T.,  2010,
  Nature, 468, 1080

\bibitem[\protect\citeauthoryear{{Meadows} \& {Seager}}{{Meadows} \&
  {Seager}}{2010}]{Meadows2010}
{Meadows} V.,  {Seager} S.,  2010, {Terrestrial Planet Atmospheres and
  Biosignatures}.
pp 441--470

\bibitem[\protect\citeauthoryear{Metchev, Marois  \& Zuckerman}{Metchev
  et~al.}{2009}]{Metchev2009}
Metchev S.,  Marois C.,   Zuckerman B.,  2009, The Astrophysical Journal
  Letters, 705, L204

\bibitem[\protect\citeauthoryear{Petit dit de~la Roche, {Hoeijmakers, H. J.}
  \& {Snellen, I. A. G.}}{Petit dit de~la Roche et~al.}{2018}]{Petit2018}
Petit dit de~la Roche D. J.~M.,  {Hoeijmakers, H. J.}  {Snellen, I. A. G.}
  2018, \mn@doi [A\&A] {10.1051/0004-6361/201833384}, 616, A146

\bibitem[\protect\citeauthoryear{Rauer et~al.,}{Rauer et~al.}{2011}]{Rauer2011}
Rauer H.,  et~al., 2011, \mn@doi [A\&A] {10.1051/0004-6361/201014368}, 529, A8

\bibitem[\protect\citeauthoryear{Rebolo et~al.}{Rebolo
  et~al.}{1998}]{Rebolo1998}
Rebolo R.,  et~al., 1998, \mn@doi [Science] {10.1126/science.282.5392.1309},
  282, 1309

\bibitem[\protect\citeauthoryear{Schneider, {Dedieu, C.}, {Le Sidaner, P.},
  {Savalle, R.}  \& {Zolotukhin, I.}}{Schneider et~al.}{2011}]{Schneider2011}
Schneider J.,  {Dedieu, C.} {Le Sidaner, P.} {Savalle, R.}  {Zolotukhin, I.}
  2011, \mn@doi [A\&A] {10.1051/0004-6361/201116713}, 532, A79

\bibitem[\protect\citeauthoryear{Serabyn, Mawet  \& Burrus}{Serabyn
  et~al.}{2010}]{Serabyn2010}
Serabyn E.,  Mawet D.,   Burrus R.,  2010, \mn@doi [Nature]
  {https://doi.org/10.1038/nature09007}, 464, 1018

\bibitem[\protect\citeauthoryear{Su et~al.,}{Su et~al.}{2009}]{Su2009}
Su K. Y.~L.,  et~al., 2009, The Astrophysical Journal, 705, 314

\bibitem[\protect\citeauthoryear{Wahhaj et~al.}{Wahhaj
  et~al.}{2011}]{Wahhaj2011}
Wahhaj Z.,  et~al., 2011, \mn@doi [The Astrophysical Journal]
  {10.1088/0004-637x/729/2/139}, 729, 139

\bibitem[\protect\citeauthoryear{Wang et~al.}{Wang et~al.}{2018}]{Wang2018b}
Wang J.~J.,  et~al., 2018, \mn@doi [The Astronomical Journal]
  {10.3847/1538-3881/aae150}, 156, 192

\bibitem[\protect\citeauthoryear{Zuckerman, Rhee, Song  \& Bessell}{Zuckerman
  et~al.}{2011}]{Zuckerman2011}
Zuckerman B.,  Rhee J.~H.,  Song I.,   Bessell M.~S.,  2011, \mn@doi [The
  Astrophysical Journal] {10.1088/0004-637x/732/2/61}, 732, 61

\bibitem[\protect\citeauthoryear{Zurlo et~al.}{Zurlo et~al.}{2016}]{Zurlo2016}
Zurlo A.,  et~al., 2016, \mn@doi [A\&A] {10.1051/0004-6361/201526835}, 587, A57

\makeatother
\end{thebibliography}








\bsp	
\label{lastpage}
\end{document}